%% file: Draft_etapri_jpsi.tex
\documentclass[aps,amsmath,showpacs,superscriptaddress,twocolumn,reprint,tightenlines]{revtex4}
\usepackage{graphicx}
\usepackage{epsfig}
\usepackage{dcolumn}
\usepackage{bm}
\usepackage{overpic}
\usepackage{color}

\newcommand{\EE}{e^+e^-}

\newcommand{\dcone}{e^{+}e^{-}\to\eta^{\prime} J/\psi}
\newcommand{\dctwo}{J/\psi \to e^{+}e^{-}}
\newcommand{\dcthree}{J/\psi \to \mu^{+}\mu^{-}}
\newcommand{\dcfour}{\eta^{\prime} \to \gamma \pi^{+}\pi^{-}}
\newcommand{\dcfive}{\eta^{\prime} \to \eta \pi^{+}\pi^{-}}

\begin{document}

\parskip=5pt plus 1pt minus 1pt

\title{\boldmath Observation of $e^{+}e^{-} \to \eta^{\prime} J/\psi$ at center-of-mass energies between 4.189 and 4.600~GeV}

\input{authors_PUB}

\date{\today}

\begin{abstract}
The process $e^{+}e^{-}\to \eta^{\prime} J/\psi$  is observed for
the first time with a statistical significance of $8.6\sigma$ at
center-of-mass energy $\sqrt{s} = 4.226$~GeV and $7.3\sigma$ at $\sqrt{s} = 4.258$~GeV
using data samples collected with the BESIII detector. The Born
cross sections are measured to be $(3.7 \pm 0.7 \pm 0.3)$ and
$(3.9 \pm 0.8 \pm 0.3)$~pb at $\sqrt{s} = 4.226$ and $4.258$~GeV,
respectively, where the first errors are statistical and the
second systematic. Upper limits at the 90\% confidence level of
the Born cross sections are also reported at other 12
energy points.
\end{abstract}

\maketitle

\section{Introduction}

The region of center-of-mass (c.m.) energies above the open charm
threshold is of great interest due to the richness of
charmonium states, whose properties are not well understood. Until
now, the vector states $\psi(3770)$, $\psi(4040)$, $\psi(4160)$, and
$\psi(4415)$ are well established experimentally in the hadronic
cross section in $\EE$ annihilation~\cite{PDG} and match very well
with the calculation in the quark model of charmonium~\cite{PM}.
By exploiting the initial state radiation (ISR) process, the B-factories BaBar
and Belle discovered several new charmonium-like
vector states, the $Y(4260)$, $Y(4360)$, and $Y(4660)$, via their decays
into the hidden-charm final states $\pi^+\pi^- J/\psi$ or $\pi^+\pi^-
\psi(3686)$~\cite{Y(4260)1, Y(4360)1, Y(4260)2,
Y(4360)2, Y(4260)3}, while there are no corresponding
structures observed in the cross sections to open-charm or inclusive
hadronic final states.  In contrast, the decay of the excited $\psi$ states
into the above two hidden-charm final states has not been observed to date.
The overpopulation of
the vector states between 4.0 and 4.7~$\rm{GeV}/\emph{c}^2$ triggered many
discussions about the nature of these states and the possible discovery
of new kinds of hadrons~\cite{epjc_review}.

Besides the $\pi^+\pi^-$ hadronic transitions, information on
other hadronic transitions will provide further insight on the
internal structure of these charmonium and charmonium-like states.
CLEO-c, BESIII, and Belle measured the cross section of $e^+e^-\to
\eta J/\psi$~\cite{Coan:2006rv, Ablikim:2012ht, Wang:2012bgc},
which has significant contribution from the $\psi(4040)$ and
$\psi(4160)$ decays and is different from the prediction in
Ref.~\cite{Wang:2011yh}, which is obtained by considering virtual charmed meson
loops. Treating $\eta$ and $\eta^{\prime}$ with the Light-Cone
approach and $J/\psi$ with non-relativistic QCD, and together with the
contribution of the resonance decays, the authors of
Ref.~\cite{qiao:2014} can reproduce the measured $e^+e^-\to \eta
J/\psi$ line shape and predict the production cross section of the
analogous process $e^{+}e^{-}\to \eta^{\prime}J/\psi$ at c.m. energies $\sqrt{s}$ from 4.3 to 5.3~GeV.

To check the theoretical predictions~\cite{qiao:2014} and to search for
potential $\eta^{\prime}J/\psi$ transitions from charmonium and
charmonium-like states, we measure the process $e^+e^-\to \eta^{\prime}
J/\psi$ with the data taken at BESIII. The CLEO-c experiment
searched for this process with data at c.m. energies $\sqrt{s}$ from
3.970 to 4.260~GeV and did not observe the signal~\cite{Coan:2006rv}.

In this paper, we report measurements of the Born cross section for
$\dcone$ at 14 energy points $\sqrt{s}$ from $4.189$ to
$4.600$~GeV~\cite{ecm_gaoq}. The data samples are collected with
the BESIII detector~\cite{ref:bes3} operating at the BEPCII storage
ring. The total integrated luminosity is about 4.5~$\rm{fb^{-1}}$,
which is measured using large angle Bhabha events with an
uncertainty of 1\%~\cite{ref:luminosity}. In the analysis, the
$J/\psi$ is reconstructed through its decays into lepton pairs
$J/\psi\to \ell^+\ell^-$ ($\ell=e$ or $\mu$), while the $\eta^{\prime}$
is reconstructed in two decay channels, $\eta^{\prime}\to
\eta\pi^{+}\pi^{-}$ (with $\eta\to \gamma\gamma$) and
$\eta^{\prime}\to \gamma\pi^{+}\pi^{-}$.

\section{Detector and Monte Carlo simulation}

The BESIII~\cite{ref:bes3} detector is a general purpose spectrometer
at the BEPCII accelerator~\cite{ref:bes2} for studies of hadron
spectroscopy and physics in the $\tau$-charm energy
region~\cite{ref:bes3physics}. The design peak luminosity of the
double-ring $e^+e^-$ collider, BEPCII, is $10^{33}$
cm$^{-2}\rm{s^{-1}}$ at $\sqrt{s}=3.77$\,GeV with a beam current of 0.93\,A.

The BESIII detector with a geometrical acceptance of 93\% of
4$\pi$ consists of the following main components: 1) a main drift
chamber (MDC) equipped with 6796 signal wires and 21884 field
wires arranged in a small cell configuration with 43 layers
working in a gas mixture of He (40\%) and $\rm{C}_3\rm{H}_8$ (60\%).
The single wire resolution on average is 135\,$\mu$m, and the
momentum resolution for charged particles in a 1\,T magnetic field
is 0.5\% at 1\,GeV;
2) a time-of-flight system
(TOF) for particle identification made of 176 pieces of 5\,cm
thick, 2.4\,m long plastic scintillators arranged as a cylinder
with two layers for the barrel, and 96 fan-shaped, 5\,cm thick,
plastic scintillators for two end-caps. The time resolution is
80\,ps in the barrel, and 110\,ps in the end-caps, corresponding
to a K/$\pi$ separation at $2\sigma$ level up to about 1.0\,GeV;
3) an electromagnetic calorimeter (EMC) made
of 6240 CsI(Tl) crystals arranged in a cylindrical shape,
complemented by two endcaps. The energy resolution is 2.5\% in the
barrel and 5\% in the endcaps at 1.0\,GeV; the position resolution
is 6\,mm in the barrel and 9\,mm in the endcaps at 1.0\,GeV. The
time resolution of the EMC is 50\,ns.
4) a muon chamber system (MUC) in the iron flux return yoke of the
solenoid, made of resistive plate chambers
(RPC) arranged in 9 layers in the barrel and 8 layers in the
endcaps, with a resolution of 2\,cm.

In order to optimize the selection criteria, determine the
detection efficiency and estimate potential background contributions,
Monte Carlo (MC) simulated data samples are generated using a {\sc
geant4}-based~\cite{geant4} software, which takes into account the
detector geometry and material description, the detector response
and signal digitization, as well as the records of the detector
running conditions and performances. The signal MC samples of
$e^{+}e^{-} \to \eta^{\prime} J\psi$ are generated at each c.m.
energy point assuming that the Born cross section follows an incoherent
sum of a Breit-Wigner (BW) function for the $\psi(4160)$ resonance
and a polynomial term for the continuum production. For the
background study, inclusive MC samples including the $Y(4260)$
decays, ISR production of the vector charmonium states, continuum
production of hadrons and QED processes are generated with {\sc
kkmc}~\cite{kkmc,event} at $\sqrt{s}= 4.258$, 4.416, and
4.600\,GeV. For the inclusive MC samples, the main known decay
modes are generated with {\sc evtgen}~\cite{event}, and the
remaining events associated with charmonium decays are generated
with the {\sc lundcharm}~\cite{lundcharm} model, while continuum
hadronic events are generated with {\sc pythia}~\cite{ref:PYTHIA}.

\section{Event Selection and Study of Background Shape}

The candidate events of $e^{+}e^{-} \to \eta^{\prime} J/\psi$ are
required to have four charged tracks with zero net charge. All
charged tracks are required to be well reconstructed in the MDC
with good helix fit quality and to satisfy
$|\rm{cos}\theta|<0.93$, where $\theta$ is the polar angle of the
track in the laboratory frame. The charged tracks are required to
originate from the interaction region with $R_{xy}<1.0$ cm and
$|R_z|<10.0$ cm, where $R_{xy}$ and $R_z$ are the distances of closest approach
of the charged track to the interaction point perpendicular to and along the beam
direction, respectively. A charged track with momentum less than
0.8\,GeV is assigned to be a pion candidate, while a track with
momentum larger than 1.0\,GeV is assigned to be a lepton
candidate. Electron and muon separation is carried out by the
ratio $E/p$ of energy deposited in the EMC and momentum measured
in the MDC. For electron candidates, we require an $E/p$ ratio larger
than 0.8, while for muon candidates, the $E/p$ ratio is required
to be less than 0.4.

Photon candidates are reconstructed from showers in the EMC crystals.
The minimum energy of photon is required to be 25\,MeV in the
barrel ($|\cos\theta|<0.80$) or 50\,MeV in the end-cap ($0.86 <
|\cos\theta| < 0.92$). To eliminate showers produced by charged
particles, the angle between the shower and the nearest charged track
is required to be greater than 20 degrees. EMC cluster timing is
further required to be between 0 and 700\,ns
to suppress electronic noise and energy deposits unrelated
to the event. The number of good photon candidates is required to be at
least 1 for $\eta^{\prime}\to \gamma\pi^{+}\pi^{-}$ and at least 2
for $\eta^{\prime}\to \eta\pi^{+}\pi^{-}$.

For $\dcfour$, a four-constraint (4C) kinematic fit is performed
on the four selected charged tracks ($\pi^{+}\pi^{-}e^{+}e^{-}$ or
$\pi^{+}\pi^{-}\mu^{+}\mu^{-}$) and one good photon candidate to
improve the momentum and energy resolutions of the final-state
particles and to reduce the potential background. If there is more
than one photon in an event, the one resulting in the minimum $\chi^{2}_{\rm
4C}$ of the kinematic fit is retained for further study. The
$\chi^{2}_{\rm 4C}$ is required to be less than 40. For $\dcfive$,
a five-constraint (5C) kinematic fit is performed on the four charged
tracks ($\pi^{+}\pi^{-}e^{+}e^{-}$ or
$\pi^{+}\pi^{-}\mu^{+}\mu^{-}$) and two good photon candidates,
with the additional constraint on the invariant mass of $\gamma\gamma$ to be
equal to the $\eta$ nominal mass~\cite{PDG}. For events with
more than two photons, the combination with the minimum
$\chi^{2}_{\rm 5C}$ is chosen. The $\chi_{\rm 5C}^2$ is required
to be less than 40.

Besides the requirements described above, the following selection
criteria are applied to select the signal. For the decay channel
$\dcfour$, in order to eliminate the backgrounds from ISR processes with
$\psi(3686)$ in the final state or from the process $e^{+}e^{-} \to
\pi^{+}\pi^{-}J/\psi$ with Final State Radiation (FSR) from
the leptons, the invariant mass of $\pi^{+}\pi^{-}J/\psi$
($M(\pi^{+}\pi^{-}J/\psi)$) and the invariant mass of the system recoiling against
$\pi^{+}\pi^{-}$ ($M^{\rm recoil}(\pi^{+}\pi^{-})$) are required
to be out of the regions $3.65 <M(\pi^{+}\pi^{-}J/\psi)<
3.71~\rm{GeV}/\emph{c}^2$ and $3.05 <M^{\rm recoil}(\pi^{+}\pi^{-})<
3.15~\rm{GeV}/\emph{c}^2$, respectively. For the decay channel $\dcfive$, the corresponding
distributions are required to be out of the regions $3.67
<M(\pi^{+}\pi^{-}J/\psi)< 3.71~\rm{GeV}/\emph{c}^2$ and $3.65 <M^{\rm
recoil}(\pi^{+}\pi^{-})< 3.69~\rm{GeV}/\emph{c}^2$ to eliminate the
background reactions $e^{+}e^{-}\to \eta\psi(3686)\to \eta
\pi^{+}\pi^{-}J/\psi$ and $e^{+}e^{-} \to
\pi^{+}\pi^{-}\psi(3686)\to \pi^{+}\pi^{-}\eta J/\psi$,
respectively.

\begin{figure}[htbp]
\begin{center}
\begin{overpic}[height=5.0cm,angle=0]{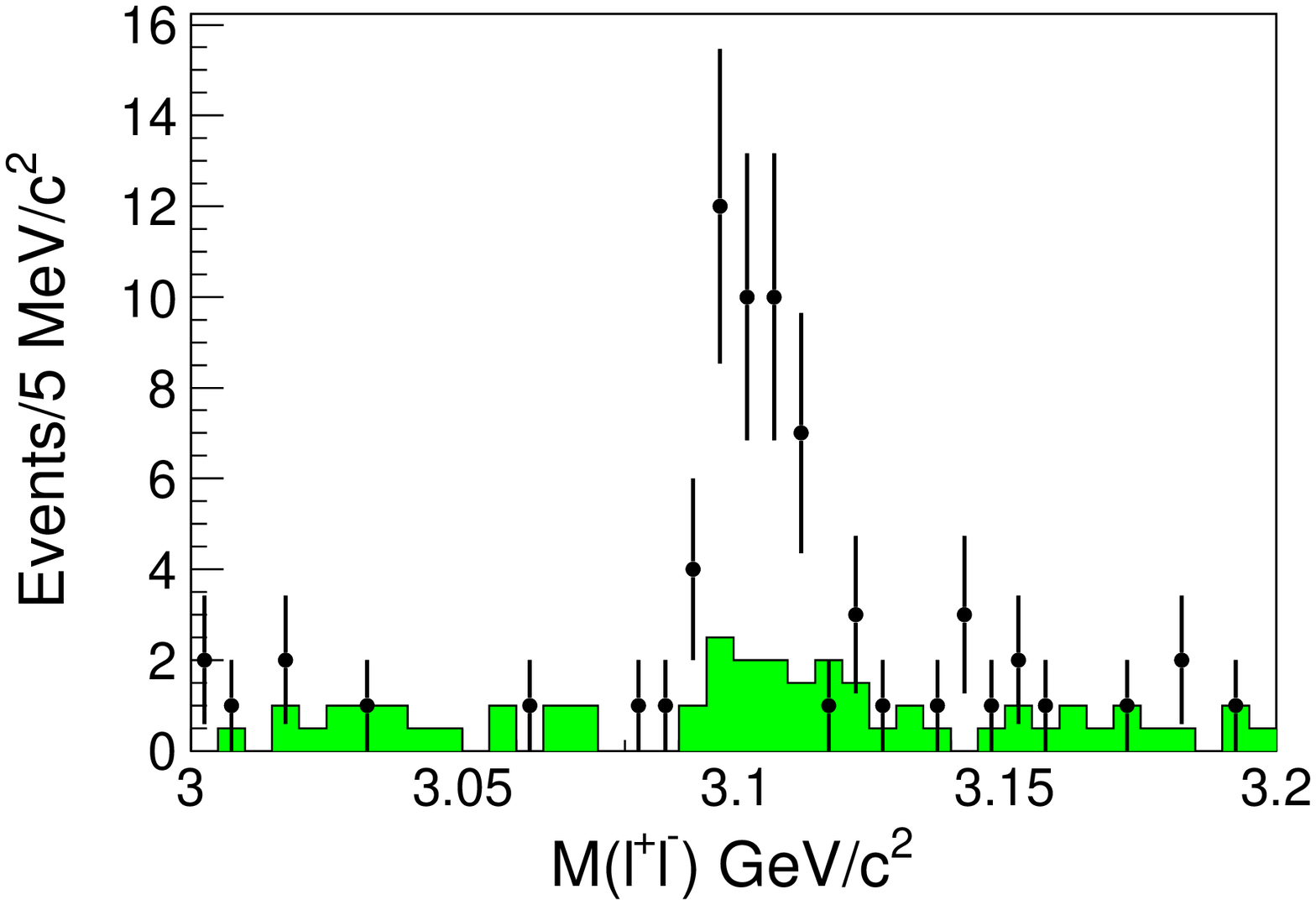}
\put(21,57){\large (a)}
\end{overpic}
\begin{overpic}[height=5.0cm,angle=0]{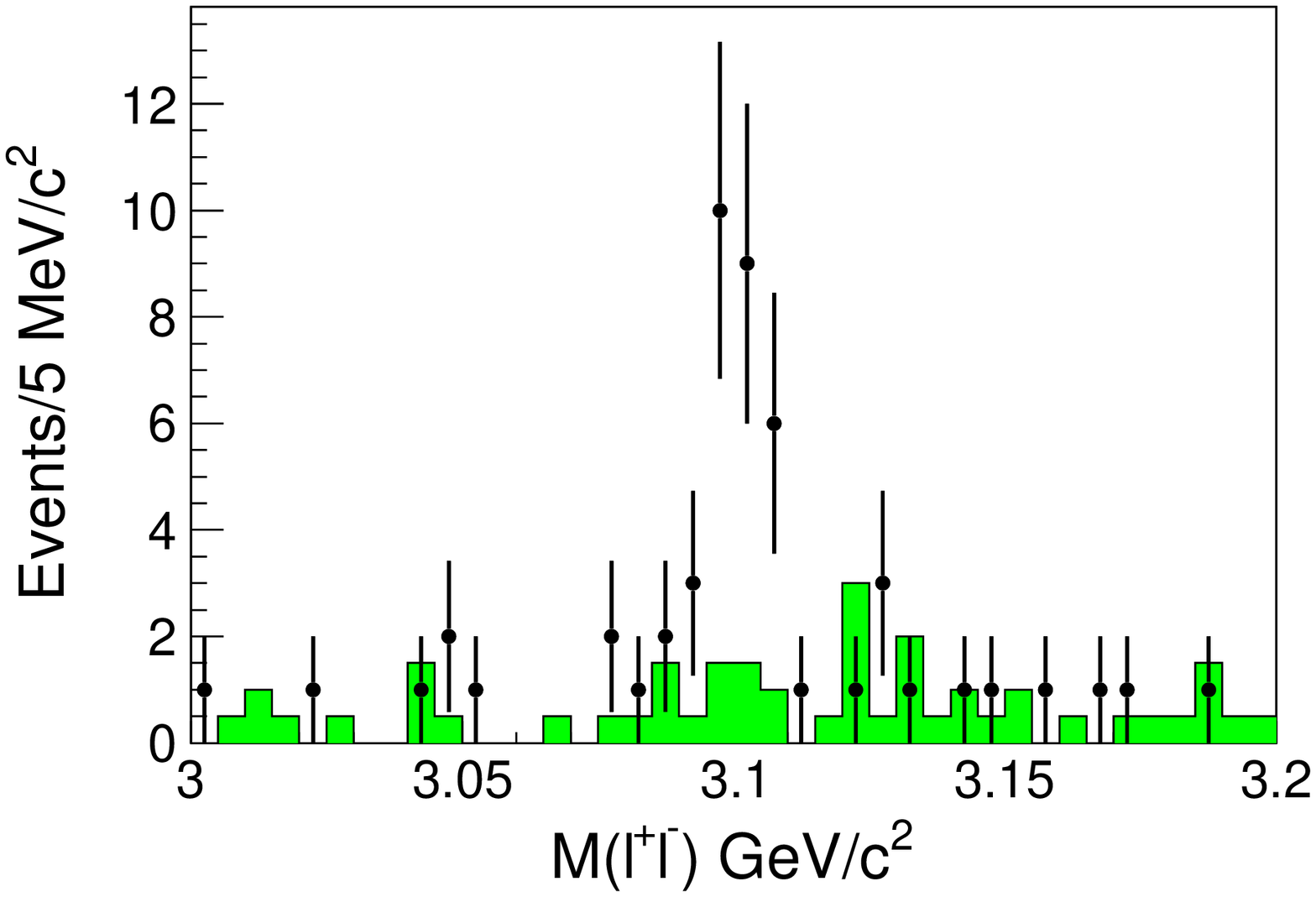}
\put(21,57){\large (b)}
\end{overpic}
\end{center}
\caption{The $M(\ell^{+}\ell^{-})$ distribution of data summed
over the four channels ($\eta^{\prime} \to
\eta\pi^{+}\pi^{-}/\gamma\pi^{+}\pi^{-}$ and $J/\psi \to
\EE/\mu^+\mu^-$) at (a) $\sqrt{s}$ = 4.226\,GeV and (b)
$\sqrt{s}$ = 4.258\,GeV. The dots with error bars and the (green) shaded
histograms represent events within $\eta^{\prime}$ signal and
sideband regions, respectively.} \label{jpsi}
\end{figure}

After applying the above selection criteria, Fig.~\ref{jpsi} shows
the invariant mass distribution of $\ell^{+}\ell^{-}$ for events
with the invariant mass of $\gamma(\eta) \pi^{+}\pi^{-}$ within
the $\eta^{\prime}$ signal and sideband regions for the data
samples at $\sqrt{s}$ = 4.226 and 4.258\,GeV. Here, the
$\eta^{\prime}$ signal region is defined as ($0.94, 0.98) ~\rm{GeV}/\emph{c}^2$,
while $\eta^{\prime}$ sideband regions are ($0.90, 0.94)~\rm{GeV}/\emph{c}^2$ and
($0.98, 1.02)~\rm{GeV}/\emph{c}^2$. The $J/\psi$ signals are observed clearly at
both energy points.  According to the MC study, the small peaking background visible in
the sideband distribution around the $J/\psi$ mass comes from $e^{+}e^{-}\to \gamma_{\rm ISR}
\pi^{+}\pi^{-}J/\psi$, which does not
produce peaking background in the distribution of
$M(\gamma\pi^{+}\pi^{-})$. The mass window requirement $3.07
<M(\ell^{+}\ell^{-})< 3.13~\rm{GeV}/\emph{c}^2$ is used to select $J/\psi$
signal for further study. After imposing all these selection
criteria, the background contribution is investigated with the
inclusive MC samples. The dominant backgrounds are found to be
those with the same final states as the signal events but without
$\eta^{\prime}$ or $J/\psi$ intermediate states, and can not be
eliminated completely.

\section{\bf Signal Determination}

\begin{figure}[htbp]
\begin{center}
\begin{overpic}[height=6cm,angle=0]{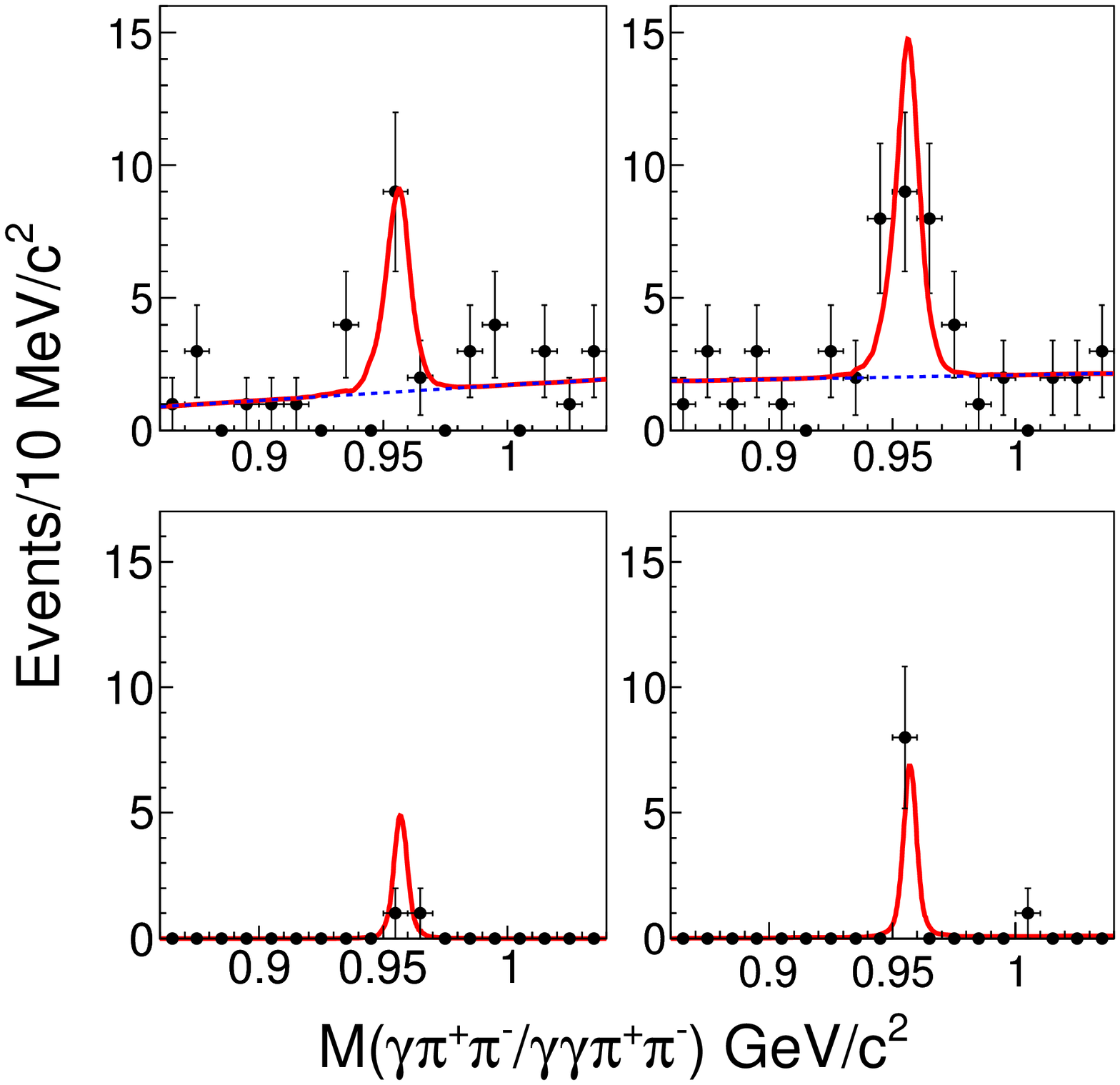}
\put(17,82){(a)}
\put(62,82){(b)}
\put(17,41){(c)}
\put(62,41){(d)}
\end{overpic}
\begin{overpic}[height=6cm,angle=0]{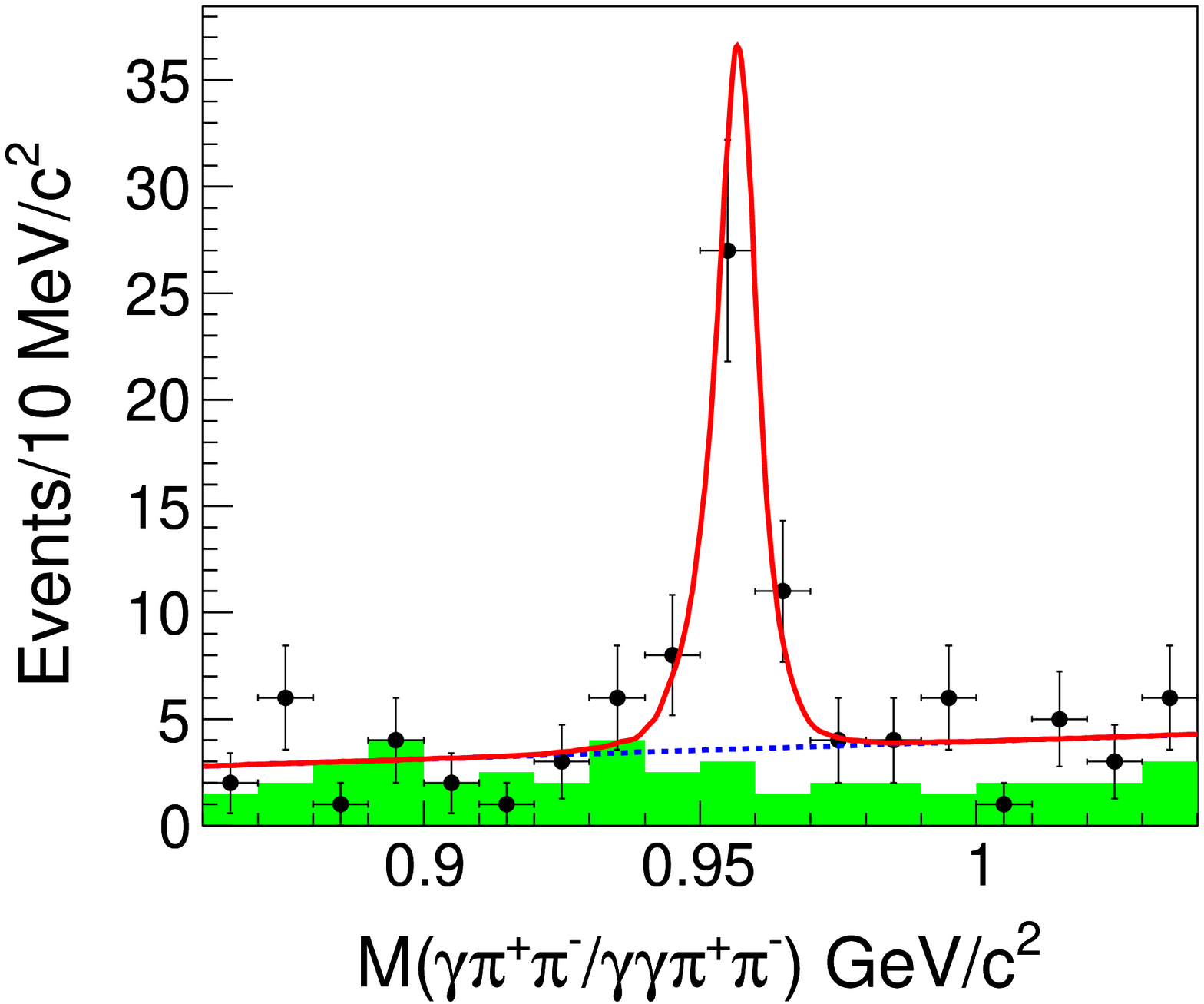}
\put(22,68){(e)}
\end{overpic}
\end{center}
\caption{Simultaneous fit to the
$M(\gamma\pi^{+}\pi^{-}/\gamma\gamma\pi^{+}\pi^{-})$ spectra at
$\sqrt{s}$ = 4.226\,GeV. (a) for $\dcfour$ and $\dctwo$, (b) for
$\dcfour$ and $\dcthree$, (c) for $\dcfive$ and $\dctwo$, (d) for
$\dcfive$ and $\dcthree$. (e) shows the combined result. The dots
with error bars and the (green) shaded histograms represent events
from data within the $J/\psi$ signal and sideband regions,
respectively. The solid lines show the fit results, while the
dashed lines represent the background. } \label{fit1}
\end{figure}

\begin{figure}[htbp]
\begin{center}
\begin{overpic}[height=6cm,angle=0]{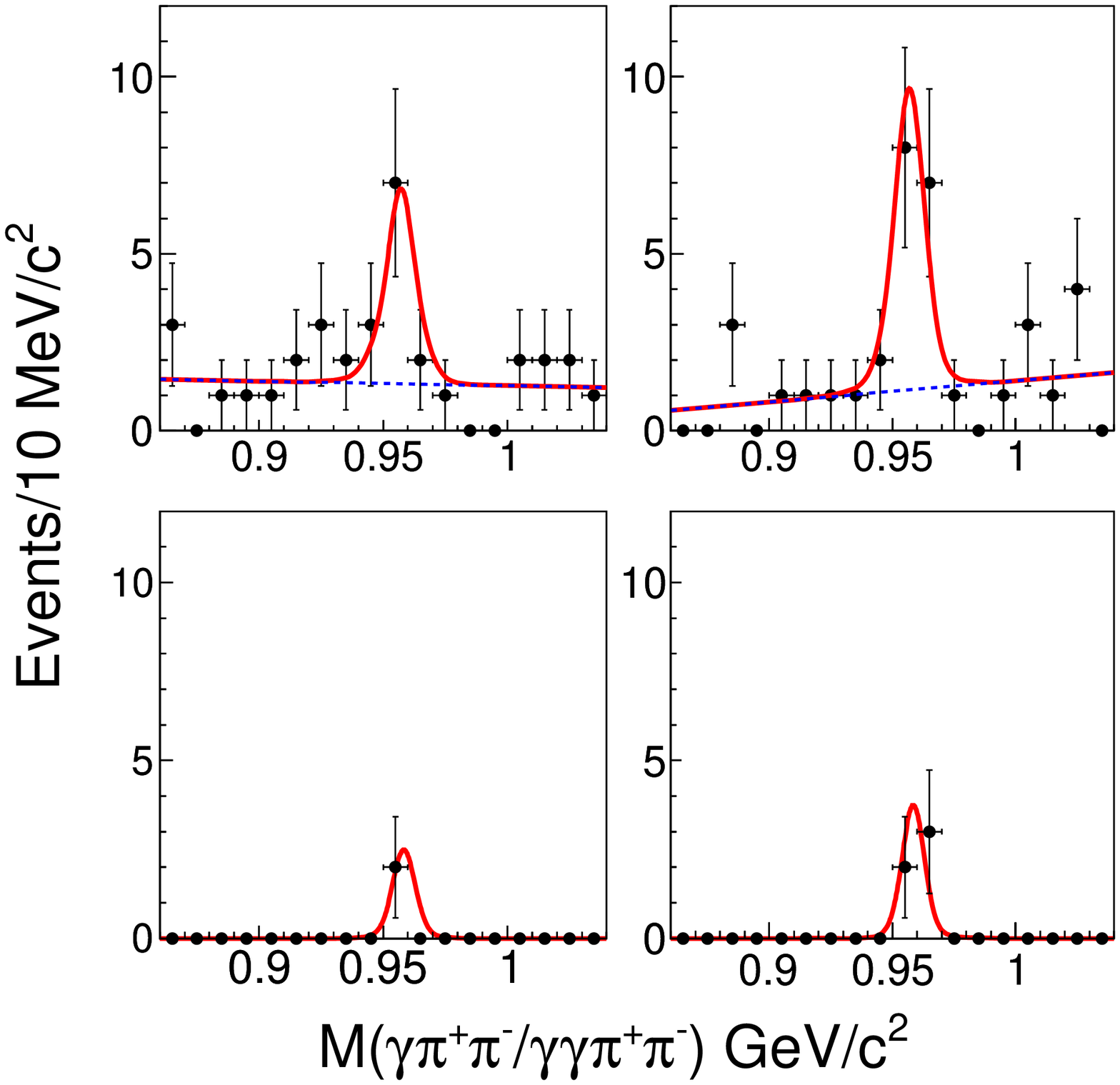}
\put(17,82){(a)}
\put(62,82){(b)}
\put(17,41){(c)}
\put(62,41){(d)}
\end{overpic}
\begin{overpic}[height=6cm,angle=0]{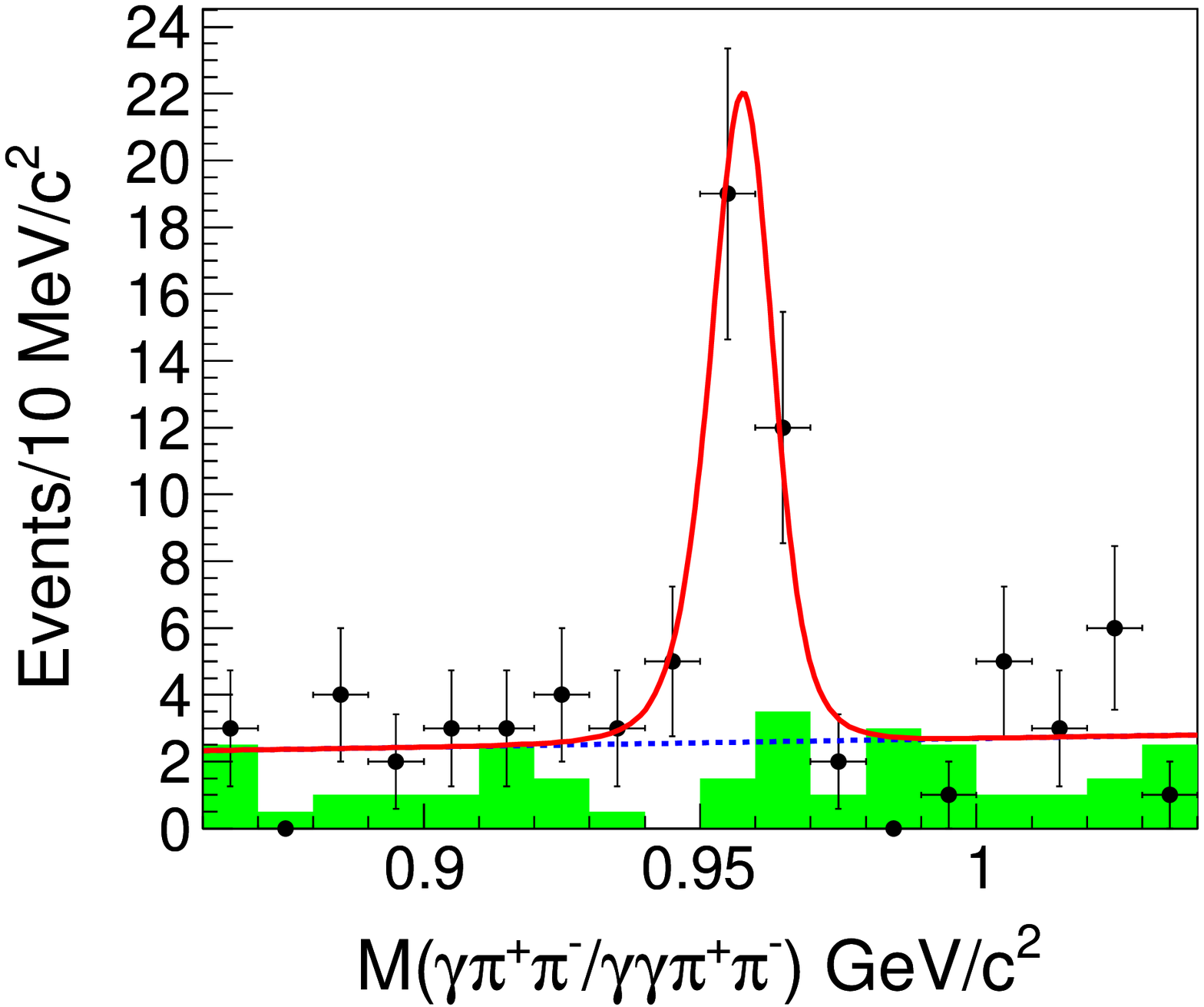}
\put(22,68){(e)}
\end{overpic}
\end{center}
\caption{Simultaneous fit to the
$M(\gamma\pi^{+}\pi^{-}/\gamma\gamma\pi^{+}\pi^{-})$ spectra at
$\sqrt{s}$ = 4.258\,GeV. (a) for $\dcfour$ and $\dctwo$, (b) for
$\dcfour$ and $\dcthree$, (c) for $\dcfive$ and $\dctwo$, (d) for
$\dcfive$ and $\dcthree$. (e) shows the combined result. The dots
with error bars and the (green) shaded histograms represent events
from data within the $J/\psi$ signal and sideband regions,
respectively. The solid lines show the fit results, while the
dashed lines represent the background.}
\label{fit2}
\end{figure}

After applying all of the above selection criteria except for the
$\eta^{\prime}$ mass window requirement, the invariant mass
distributions of $\gamma\pi^{+}\pi^{-}$ and $\eta\pi^{+}\pi^{-}$
for $J/\psi\to e^{+}e^{-}$ and $J/\psi \to \mu^{+}\mu^{-}$ individually
as well as the combination of four channels are
shown in Fig.~\ref{fit1} and Fig.~\ref{fit2} for the data at $\sqrt{s}=4.226$
and 4.258\,GeV, respectively. The $\eta^{\prime}$ is observed clearly in the combined
distribution.
The background is a flat distribution in the $\gamma\pi^{+}\pi^{-}$ invariant mass; this
is verified by studying the corresponding
distributions of the events in the $J/\psi$ sideband region and of
the MC samples. The invariant mass distribution of the
$\eta\pi^{+}\pi^{-}$ channel is essentially background free.

To determine the signal yields, a simultaneous fit to the invariant
mass of $\gamma (\eta) \pi^{+}\pi^{-}$ with an unbinned maximum
likelihood method is performed for the four different channels.
The total signal yield, denoted as $N^{\rm tot}$, is
a free parameter in the fit. The signal yields for the individual decay modes are
constrained by assuming the same production cross section for
$e^{+}e^{-} \to \eta^{\prime} J/\psi$ and are determined to be
$N^{\rm tot} \times {\cal B} (\eta^{\prime}) \times {\cal B}
(J/\psi) \times \epsilon$, where ${\cal B} (\eta^{\prime})$ and
${\cal B} (J/\psi)$ are the decay branching fractions of
$\eta^{\prime}$ and $J/\psi$, respectively, and $\epsilon$ is the
corresponding detection efficiency. The $\eta^{\prime}$ signal is
described by a MC simulated shape convolved with a Gaussian
function to take into account the mass resolution difference
between data and the MC simulation; the parameters of the Gaussian
function are free but constrained to be the same for the different
channels. The background is described with a linear function, and
its normalization factors are allowed to vary in different
channels.

Projections of the mode-by-mode and combined fit results at $\sqrt{s}$ =
4.226\,GeV are shown in Fig.~\ref{fit1}. The $\chi^2 /{\rm ndf}$
for the combined result is 0.9, where sparsely populated bins are
combined so that there are at least seven counts per bin in the
$\chi^2$ calculation and ${\rm ndf}$ is the number of degrees of
freedom. The fit yields $N^{\rm obs}= 36.5\pm 6.9$, and the
statistical significance of the $\eta^{\prime}$ signal is
determined to be $8.6\sigma$ by comparing the log-likelihood values
with and without $\eta^{\prime}$ signal included in the fit and taking the
change of the number of free parameters into account. A similar fit process is
performed for the data at $\sqrt{s}$ = 4.258\,GeV, and
corresponding results are shown in Fig~\ref{fit2}. The $\chi^2
/{\rm ndf}$ for the combined result is 0.94, the fit yields
$N^{\rm obs} = 30.0 \pm 6.2$ and the statistical significance of
the $\eta^{\prime}$ signal is $7.3\sigma$.

\begin{figure*}[htbp]
\begin{center}
\begin{overpic}[height=6.5cm,angle=0]{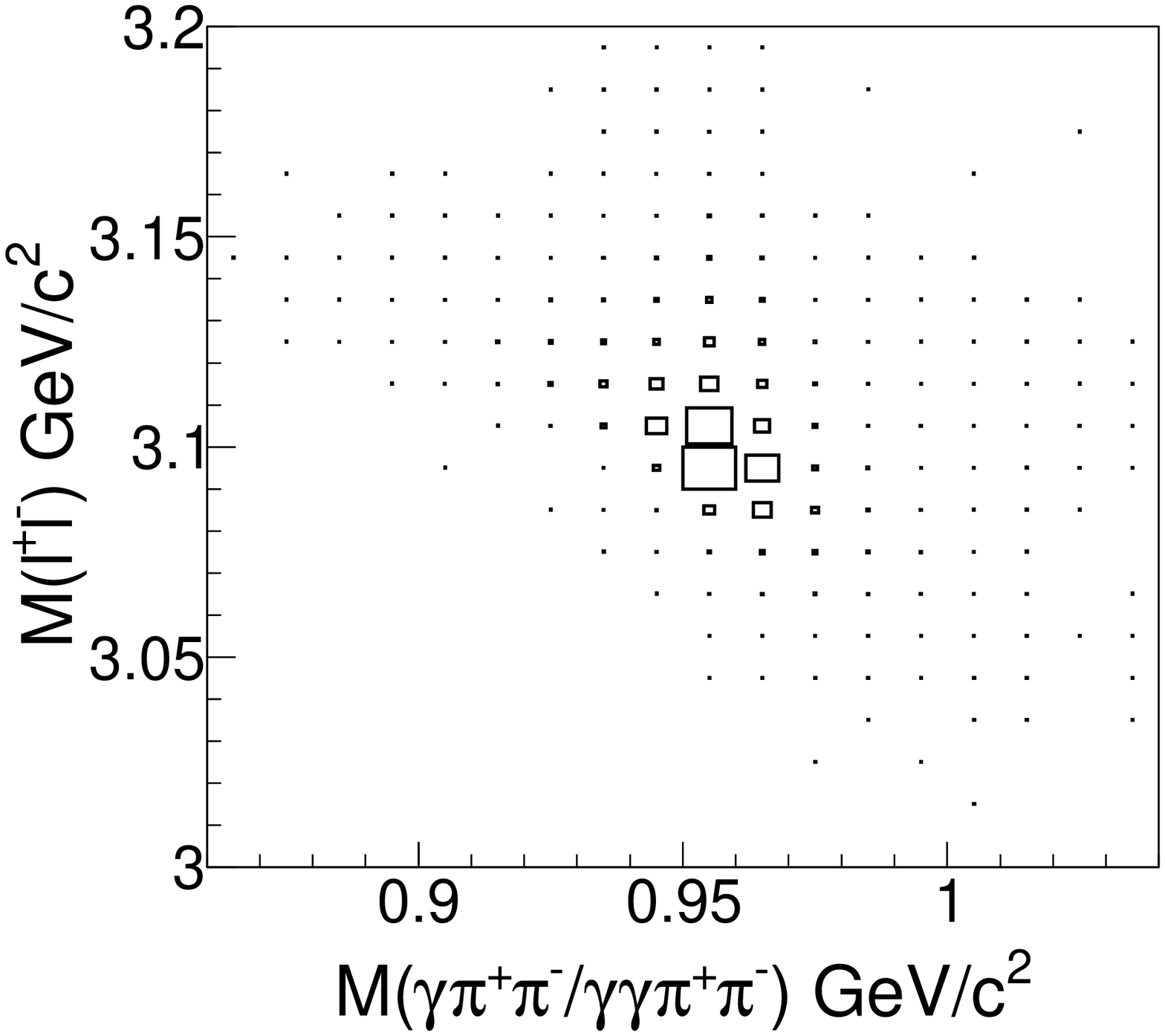}
\put(20,76){(a)}
\end{overpic}
\begin{overpic}[height=6.5cm,angle=0]{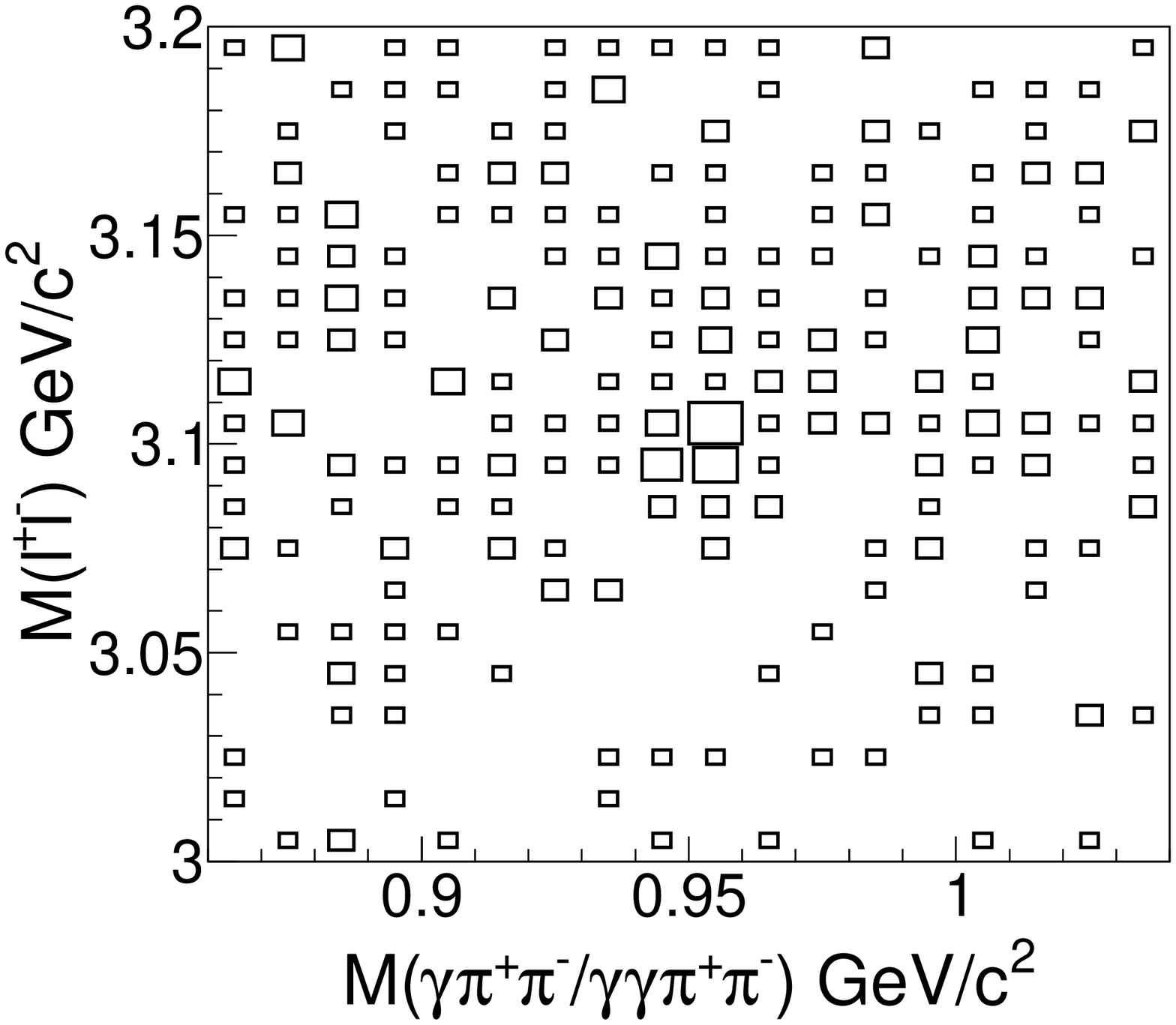}
\put(20,75){(b)}
\end{overpic}
\begin{overpic}[height=6.1cm,angle=0]{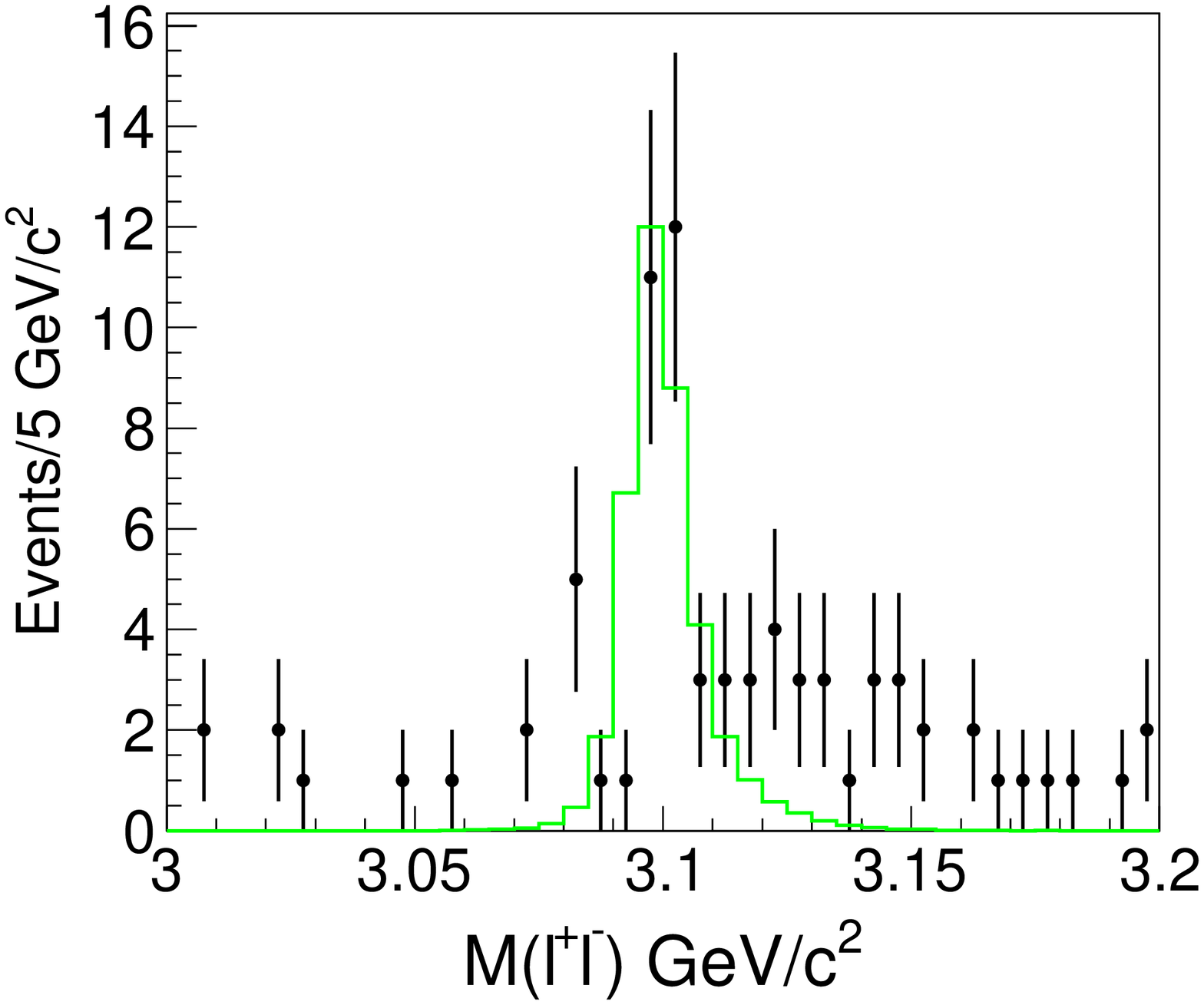}
\put(20,72){(c)}
\end{overpic}
\begin{overpic}[height=6.1cm,angle=0]{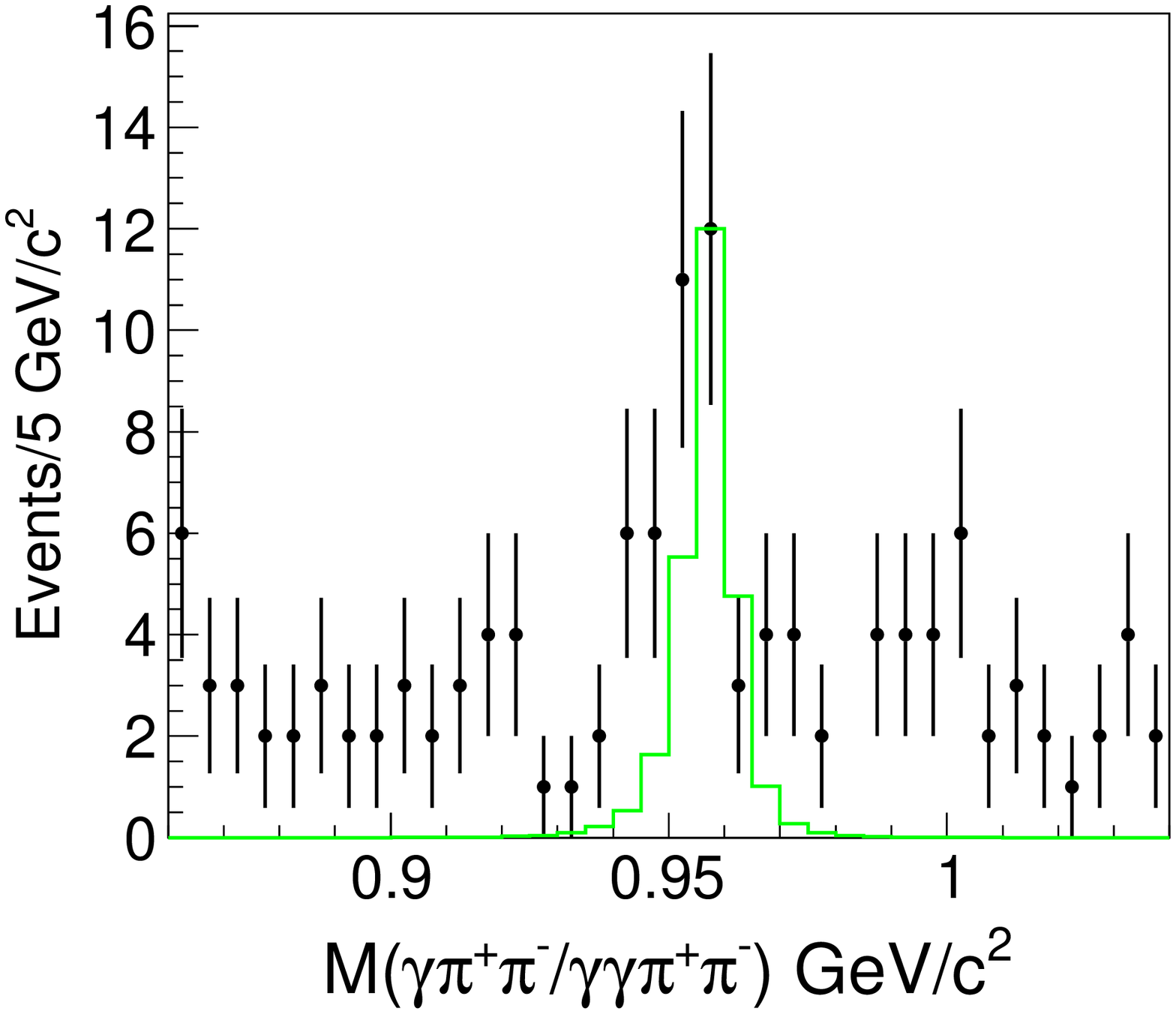}
\put(20,72){(d)}
\end{overpic}
\end{center}
\caption{The distributions for the data samples taken at
$\sqrt{s}$ = 4.189, 4.208, 4.217, 4.242, 4.308, 4.358, 4.387,
4.416, 4.467, 4.527, 4.575, and 4.600\,GeV, (a) the scatter plot
of  $M(\ell^{+}\ell^{-})$ versus
$M(\gamma\pi^{+}\pi^{-}/\eta\pi^{+}\pi^{-})$ for the MC
simulation; (b) the corresponding scatter plot for the data; (c)
the projection of $M(\ell^{+}\ell^{-})$, and (d) the projection of
$M(\gamma\pi^{+}\pi^{-}/\eta\pi^{+}\pi^{-})$, in which points with
error bars are data and histograms are signal MC
simulation.}\label{2D}
\end{figure*}

The same event selection criteria are applied to the data samples
taken at the other $12$ energy points. Figure~\ref{2D} depicts the scatter
plot of $M(\ell^{+}\ell^{-})$ versus
$M(\gamma\pi^{+}\pi^{-}/\eta\pi^{+}\pi^{-})$ and the projections
of $M(\ell^{+}\ell^{-})$ and
$M(\gamma\pi^{+}\pi^{-}/\eta\pi^{+}\pi^{-})$ including all 12
energy points. We can see a cluster of events in the signal
region, although no significant $\eta^{\prime}J/\psi$ signal is
observed at any individual energy point. As a consequence, upper
limits on the number of signal events at the $90\%$ confidence
level (C.L.) are set using a Bayesian method~\cite{bayes} at
every individual energy point. By fitting the
  $M(\gamma\pi^{+}\pi^{-}/\eta \pi^{+}\pi^{-})$ distribution with fixed values for the
  signal yield, we obtain a scan of the likelihood as a
  function of the number of signal events. The upper limit is determined by finding the
  number of signal events below which lies 90$\%$ of the area under the likelihood
  distribution. The results are listed in Table~\ref{ul}.

\section{Cross section results}

The Born cross section is calculated with
\begin{equation}
\sigma^{\rm B} = \frac{N^{\rm obs}}{L_{\rm int}\cdot (1+\delta)
\cdot |1+\Pi|^{2} \cdot \sum_{i=1}^{4} \epsilon_{\rm i} \cal B_{\rm i}} ,
\end{equation}
where
$L_{\rm int}$ is the integrated luminosity, $\epsilon_{\rm i}$ is selection efficiency for the $i$th channel estimated from
the MC simulation, $\cal {B}_{\rm i}$ is the product branching
fraction of the intermediate states for the $i$th channel taken from
the Particle Data Group~\cite{PDG}, $|1+\Pi|^{2}$ is the vacuum
polarization factor~\cite{VP} and $(1+\delta)$ is the radiative
correction factor, which is defined as
\begin{equation}
    1+\delta = \frac{ \int^{1}_{0} \sigma(s(1-x)) F(x,s)dx}{\sigma(s)}.
\end{equation}
The radiative correction changes the total cross section, and
emission of additional photons affects the efficiency of
selection. Here, $x$ is the ratio between radiative photon's
energy and the center of mass energy, $F(x,s)$ is the radiator
function, which is obtained from a QED calculation~\cite{QED} with
an accuracy of $0.1\%$, and $\sigma(s)$ is the line shape of the cross section
for $e^{+}e^{-}\to \eta^{\prime}J/\psi$, which is described
by a constant-width BW function with the parameters of the
$\psi(4160)$ plus a polynomial function.

All the numbers used in the cross section calculation are
summarized in Table~\ref{ul}. The Born cross section is measured
to be $(3.7\pm 0.7)$\,pb at 4.226\,GeV and $(3.9\pm 0.8)$\,pb at 4.258\,GeV,
where the errors are statistical. The Born cross sections and upper
limits at the other energy points are also shown in Table~\ref{ul}. In
the upper limit determination, a conservative result with a factor
$1/(1-\sigma)$ is included to take into account the effect of the
total systematic uncertainty, $\sigma$, which is described in the
next section in detail.

\begin{table*}[htbp]
  \centering
\caption{Summary of the values used to calculate the Born cross
section of $\dcone$. The upper limits are at the $90\%$ C.L. }
  \begin{tabular}{ccccccc}
  \hline
  \hline
  $\sqrt{s}$\,(GeV)  &  $N^{\rm obs}$  & $L_{\rm int}$ ($\rm{pb^{-1}}$) &  1+$\delta$    &  $\sum\epsilon_{\rm i} \cal B_{\rm i}$ ($\rm{10^{-2}}$)  & $|1+\Pi|^{2}$ & $\sigma^{\rm B}$\,(pb)   \\
  \hline
   $4.189$                          &$3.8\pm2.3$ ($< 8.7$)            &43.1            & 0.857               & 1.01                      & 1.056               & $9.7\pm5.8\pm0.6$ ($< 24$)      \\
   $4.208$                          &$2.6\pm3.2$ ($< 13.3$)           &54.6            & 0.885               & 1.04                      & 1.057               & $4.9\pm6.1\pm0.4$ ($< 27$)          \\
   $4.217$                          &$1.0\pm1.7$ ($< 6.2$)            &54.1            & 0.902               & 1.00                      & 1.057               & $1.9\pm3.3\pm0.2$ ($< 13$)         \\
   $4.226$                          &$36.5\pm6.9$                     &1047.3          & 0.919               & 0.98                      & 1.056               & $3.7\pm0.7\pm0.3$                   \\
   $4.242$                          &$0.8\pm1.4$ ($< 5.3$)            &55.6            & 0.945               & 0.95                      & 1.056               & $1.5\pm2.7\pm0.2$ ($< 11$)        \\
   $4.258$                          &$30.0\pm6.2$                     &825.7           & 0.969               & 0.91                      & 1.054               & $3.9\pm0.8\pm0.3$                   \\
   $4.308$                          &$2.2\pm1.5$ ($< 5.9$)            &44.9            & 1.036               & 0.81                      & 1.052               & $5.6\pm3.8\pm0.3$ ($< 16$)        \\
   $4.358$                          &$3.0\pm2.3$ ($< 7.9$)            &539.8           & 1.114               & 0.77                      & 1.051               & $0.6\pm0.5\pm0.1$ ($< 1.7$)         \\
   $4.387$                          &$2.1\pm2.1$ ($< 8.3$)            &55.2            & 1.162               & 0.73                      & 1.051               & $4.3\pm4.3\pm0.3$ ($< 18$)        \\
   $4.416$                          &$10.8\pm4.1$($< 15.9$)           &1028.9          & 1.191               & 0.71                      & 1.053               & $1.2\pm0.5\pm0.1$ ($< 2.0$)        \\
   $4.467$                          &$5.9\pm4.1$ ($< 14.8$)           &109.9           & 1.161               & 0.72                      & 1.055               & $6.1\pm4.2\pm0.5$($< 17$)        \\
   $4.527$                          &$1.4\pm1.3$ ($< 5.3$)            &110.0           & 1.002               & 0.81                      & 1.055               & $1.5\pm1.4\pm0.1$ ($< 6.1$)       \\
   $4.575$                          &$0.0\pm1.7$ ($< 9.0$)            &47.7            & 0.907               & 0.90                      & 1.055               & $0.0\pm4.2\pm0.4$($< 24$)          \\
   $4.600$                          &$1.2\pm2.3$ ($< 7.9$)            &566.9           & 0.880               & 0.92                      & 1.055               & $0.3\pm0.5\pm0.1$ ($< 2.1$)        \\
  \hline
    \end{tabular}
    \label{ul}
\end{table*}

\begin{figure}[htbp]
\begin{center}
\begin{overpic}[width=0.45\textwidth,angle=0]{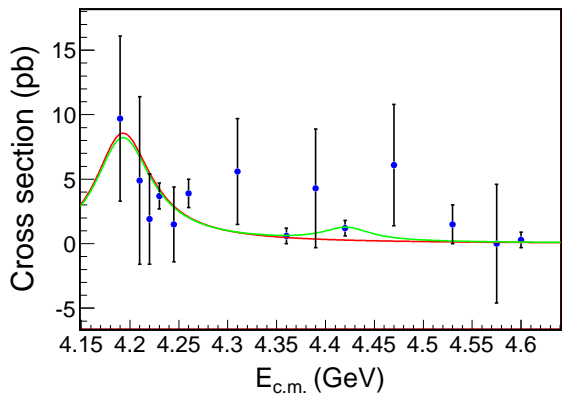}
\end{overpic}
\end{center}
\caption{Fit to the Born cross section $\sigma(e^+e^-\to
\eta^{\prime}J/\psi)$ with a $\psi(4160)$ resonance (red curve),
or a combination of $\psi(4160)$ and $\psi(4415)$ resonances
(green curve). The uncertainties are statistical only. }
\label{Fit_CS}
\end{figure}

Figure~\ref{Fit_CS} shows the measured Born cross sections for
$e^+e^-\to \eta^{\prime}J/\psi$ over the energy region studied in
this work. Assuming that the $\eta^{\prime}J/\psi$ signals come
from the $\psi(4160)$ decay, the cross section is fitted with a
constant-width relativistic BW function, i.e.,
\begin{equation}
\sigma(m) = |{\cal{A}}_{\psi(4160)}(m)\cdot \sqrt{{\Phi(m)}/{\Phi(M)}}|^{\rm 2},
\end{equation}
where ${\cal{A}}_{\psi(4160)}(m)$ represents the contribution of $\psi(4160) \to \eta^{\prime} J/\psi$ and ${\Phi(m)}$ is the 2-body phase space factor.
Here, ${\cal{A}}_{\psi(4160)}(m)$ is written as below:
\begin{equation}
{\cal{A}}_{\psi(4160)}(m) = \frac {\sqrt{12\pi\Gamma_{ee}\Gamma_{\rm tot}
\cal B(\rm \psi(4160)\to\eta^{\prime}J/\psi)}} {m^{2}-M^{2}+iM{\Gamma}_{\rm tot}},
\end{equation}
where the resonant parameters (the mass $M$, the total width $\Gamma_{\rm tot}$ and the electron partial
width $\Gamma_{ee}$) of the $\psi(4160)$
and the branching ratio for $\psi(4160) \to \eta^{\prime} J/\psi$ are taken from PDG~\cite{PDG} and fixed in the fit.
The $\chi^2 /\rm{ndf}$ is 0.9, which means the
measurement supports our assumption. The second resonance,
$\psi(4415)$~\cite{PDG}, is also added in the fit; the statistical
significance is determined to be $2.6\sigma$ by comparing the two
$-2\ln(L)$ values and taking the change of ${\rm ndf}$ into
account. It indicates that the contribution of $\psi(4415)$ is not significant.

\section{Systematic Uncertainties}

Several sources of systematic uncertainties are considered in the
measurement of the Born cross section, including the integrated
luminosity measurement, background shape, fitting range, ISR
correction factor, photon detection, tracking efficiency,
kinematic fit, lepton pair mass resolution, and the branching
fractions of intermediate states decay.

(a) The uncertainty from integrated luminosity measurement using
large angle Bhabha ($e^{+}e^{-} \to e^{+}e^{-}$) scattering is estimated to be
$1.0\%$~\cite{ref:luminosity}.

(b) The systematic uncertainty due to the background shape is
estimated by varying the background shape from a linear function
to a second order Chebyshev polynomial. The difference in the
signal yields is taken as the systematic uncertainty.

(c) The systematic uncertainty due to the fit range is estimated
by varying the fit range from $[0.86, 1.04]~\rm{GeV}/\emph{c}^2$ to $[0.87,
1.05]~\rm{GeV}/\emph{c}^2$ or $[0.85, 1.03]~\rm{GeV}/\emph{c}^2$. The
largest change in the signal yields is taken as the systematic uncertainty.

Since the relative signal yields for each individual decay mode $i$ is
constrained by the weight factor
$\epsilon_{\rm i} \cal B_{\rm i}$ / $\sum_{\rm i=1}^{4} \epsilon_{\rm i} \cal
B_{\rm i}$ in the fit procedure, the uncertainties due to $\epsilon_{\rm i}$ or
$\cal B_{\rm i}$ affect not only $\epsilon_{\rm i} \cal B_{\rm i}$
but also $N^{\rm obs}$. Taking both terms into account, we change
the values of $\epsilon_{\rm i}$ or $\cal B_{\rm i}$, then refit
the data. The change of the measured cross section is taken as the
systematic uncertainty. The following systematic uncertainties are estimated
by this method except for the tracking efficiency.
Because the four decay channels have the same, fully correlated,
uncertainty on the tracking efficiency, this uncertainty will not
affect the fit result.
Most of these uncertainties are energy independent, except that associated with ISR correction.
We use the uncertainties determined with data at the high-statistics energy point $\sqrt{s} = 4.226$\,GeV
as the systematic uncertainties for all the samples.

(d) The ISR correction factors are obtained by a QED calculation using
the cross section measured by this analysis, which is parameterized by a BW function for $\psi(4160)$ plus a polynomial function.
The ISR correction factors are calculated iteratively until they become stable.
To estimate the uncertainty due to the ISR correction factor, the measured cross section is also parameterized by a BW function
or a polynomial function. The largest discrepancy between the results with alternative assumption and
the nominal value is taken as the systematic uncertainty.

(e) The uncertainty due to photon reconstruction efficiency is
$1.0\%$ per photon~\cite{photon}. Therefore, we vary the values of
$\epsilon_{\rm i}$ up or down by $1\%\times N_{\rm \gamma}$ and
refit the data, where $N_{\rm \gamma}$ is the number of photons in
the final state. The maximum change of the measured cross section is
taken as the systematic uncertainty.

(f) The discrepancy of tracking efficiency between the MC
simulation and the data is estimated to be $1.0\%$ per charged
track from a study of $e^{+}e^{-}\to \pi^{+}\pi^{-}J/\psi$ and
$e^{+}e^{-}\to 2(\pi^{+}\pi^{-})$. There are 4 charged tracks in
the candidate events, $4.0\%$ is adopted as the changed value for
$\epsilon_{\rm i}$, so the total uncertainty in the final results is $4.0\%$.

(g) The mass resolution discrepancy between the MC simulation and
the data will introduce an uncertainty when we apply a mass window
requirement on the invariant mass distribution of the lepton pairs.
This uncertainty is estimated using the control sample
$e^{+}e^{-}\to \gamma_{\rm ISR}\psi(3686)\to \gamma_{\rm ISR}
\pi^{+}\pi^{-}J/\psi$ with $J/\psi \to e^{+}e^{-}$ or
$\mu^{+}\mu^{-}$. The same $J/\psi$ mass window $[3.07,
3.13]~\rm{GeV}/\emph{c}^2$ is required for both the data and the MC sample,
and the discrepancy in efficiency between the MC simulation and
the data is $(1.0 \pm 1.1)\%$ and $(2.9 \pm 1.6)\%$
for $J/\psi \to e^{+}e^{-}$ and $\mu^{+}\mu^{-}$, respectively.
After refitting the data, the large change on the measured cross section
with respected to the nominal value is taken as the systematic uncertainty.

(h) The uncertainty associated with the kinematic fit arises from
the inconsistency of track helix parameters between the data and
the MC simulation. Therefore, the three track parameters $\phi_{0}$,
$\kappa$, and $\tan \lambda$ are corrected in the signal MC
samples. The correction factors are obtained by comparing their
pull distributions in a control sample between data and MC
simulation~\cite{helix}. The difference of the detection
efficiency between the samples with and without the helix
correction affects the weight factors. The data is refitted and
the resulting difference on the Born cross section with respect
to the nominal value is taken as the systematic uncertainty.

(i) The branching fractions of $J/\psi \rightarrow
e^{+}e^{-}/\mu^{+}\mu^{-}$, $\eta^{\prime} \rightarrow
\gamma\pi^{+}\pi^{-}/\eta\pi^{+}\pi^{-}$, and $\eta \rightarrow
\gamma\gamma$ are changed independently.
The sum in quadrature of all individual uncertainties on the Born cross section is taken as the systematic uncertainty.

(j) Final state radiation affects both the lepton pair
invariant mass distribution and the efficiency of the kinematic fit; its systematic uncertainty is
taken into account.
The uncertainties related with the requirements to veto backgrounds are negligibly small,
and the uncertainties from other sources such as the $E/p$ ratio requirement for electron and muon separation,
the vacuum polarization and c.m.\ energy measurement are estimated to be less than 1\% and are
neglected in this analysis.

The sources of systematic uncertainty and their contributions are
summarized in Table~\ref{errors}. The total systematic uncertainty
is the sum in quadrature of all individual uncertainties.

\begin{table*}[htbp]
\scriptsize
  \centering
  \caption{Summary of systematic uncertainties ($\%$).}
  \begin{tabular}{ccccccccccccccc}
  \hline
  \hline
  Source/$\sqrt{s}$ (GeV)        &4.189    &4.208    &4.217     &4.226     &4.242     &4.258      &4.308         &4.358          &4.387          &4.415         &4.467          &4.527          & 4.575          &4.600 \\
  \hline
  Luminosity measurement         &1.0       &1.0      &1.0       &1.0       &1.0       &1.0        &1.0            &1.0            &1.0            &1.0          &1.0             &1.0            &1.0             &1.0   \\
  Background shape               &0.1       &3.4      &5.3       &0.2       &4.9       &2.4        &0.1            &0.2            &0.6            &4.6          &0.1             &0.2            &0.0             &2.9   \\
  Fit range                  &0.4       &4.1      &2.7       &2.2       &0.7       &2.2        &0.3            &5.1            &0.2            &7.7          &6.1             &1.2            &0.0             &3.5   \\
  ISR factor                     &3.0       &1.2      &3.0       &4.0       &4.4       &1.1        &2.5            &2.1            &2.9            &1.7          &1.5             &2.9            &6.0             &2.1  \\
  Photon  detection              &1.4       &1.4      &1.4       &1.4       &1.4       &1.4        &1.4            &1.4            &1.4            &1.4          &1.4             &1.4            &1.4             &1.4   \\
  Tracking  efficiency           &4.0       &4.0      &4.0       &4.0       &4.0       &4.0        &4.0            &4.0            &4.0            &4.0          &4.0             &4.0            &4.0             &4.0   \\
  Kinematic fitting              &1.1       &1.1      &1.1       &1.1       &1.1       &1.1        &1.1            &1.1            &1.1            &1.1          &1.1             &1.1            &1.1             &1.1   \\
  Lepton pair mass resolution    &2.2       &2.2      &2.2       &2.2       &2.2       &2.2        &2.2            &2.2            &2.2            &2.2          &2.2             &2.2            &2.2             &2.2   \\
  Branching fraction             &1.6       &1.6      &1.6       &1.6       &1.6       &1.6        &1.6            &1.6            &1.6            &1.6          &1.6             &1.6            &1.6             &1.6   \\
  \hline
  Total                          &6.1       &7.6      &8.5       &7.0       &8.5       &6.3        &5.8            &7.6            &6.0            &10.5         &8.2             &6.1            &8.0             &7.3  \\
  \hline
  \hline
  \end{tabular}
  \label{errors}
  \end{table*}

\section{Summary}

In summary, the process $e^{+}e^{-} \to \eta^{\prime} J/\psi$ is
investigated using data samples collected with the BESIII detector
at 14 c.m.\ energies from 4.189 to 4.600\,GeV. Significant
$e^{+}e^{-}\to \eta^{\prime}J/\psi$ signals are observed at
$\sqrt{s} = 4.226$ and $4.258$\,GeV for the first time, and the corresponding Born
cross sections are measured to be $(3.7 \pm 0.7 \pm 0.3)$ and
$(3.9 \pm 0.8 \pm 0.3)$\,pb, respectively. The upper limits of
Born cross sections at the 90\% C.L.\ are set for the other 12 c.m.\ %
energy points where no significant signal is observed. The
measured cross sections support the hypothesis that signal events of
$\eta^{\prime}J/\psi$ come from $\psi(4160)$ decays; the
contribution of $\psi(4415)$ is not evident.

Compared with the Born cross section of $e^{+}e^{-}\to \eta
J/\psi$~\cite{Ablikim:2012ht}, the measured Born cross section of
$e^{+}e^{-}\to \eta^{\prime}J/\psi$ is much smaller, which is in
contradiction to the calculation in Ref.~\cite{qiao:2014}. There
are two possible reasons contributing to this discrepancy. The
cross section of $e^{+}e^{-}\to \eta^{\prime} J/\psi$ is
investigated at an order of $O(\alpha_{s}^{4})$, therefore, higher
order correction might need to be considered; additionally,
the proportion of gluonic admixture in $\eta^{\prime}$
need to be further studied to make certain
the contribution of a gluonium component on the results.

\acknowledgments

The BESIII collaboration thanks the staff of BEPCII and the IHEP
computing center for their strong support. This work is supported in
part by National Key Basic Research Program of China under Contract
No. 2015CB856700; National Natural Science Foundation of China (NSFC)
under Contracts Nos. 11125525, 11235011, 11322544, 11335008, 11425524;
the Chinese Academy of Sciences (CAS) Large-Scale Scientific Facility
Program; the CAS Center for Excellence in Particle Physics (CCEPP);
the Collaborative Innovation Center for Particles and Interactions
(CICPI); Joint Large-Scale Scientific Facility Funds of the NSFC and
CAS under Contracts Nos. 11179007, U1232201, U1332201; CAS under
Contracts Nos. KJCX2-YW-N29, KJCX2-YW-N45; 100 Talents Program of CAS;
National 1000 Talents Program of China; INPAC and Shanghai Key
Laboratory for Particle Physics and Cosmology; German Research
Foundation DFG under Contract No. Collaborative Research Center
CRC-1044; Istituto Nazionale di Fisica Nucleare, Italy; Koninklijke
Nederlandse Akademie van Wetenschappen (KNAW) under Contract
No. 530-4CDP03; Ministry of Development of Turkey under Contract
No. DPT2006K-120470; Russian Foundation for Basic Research under
Contract No. 14-07-91152; The Swedish Resarch Council;
U. S. Department of Energy under Contracts Nos. DE-FG02-04ER41291, DE-FG02-05ER41374,
DE-SC-0010504, DE-SC0012069, DESC0010118; U.S. National Science
Foundation; University of Groningen (RuG) and the Helmholtzzentrum
fuer Schwerionenforschung GmbH (GSI), Darmstadt; WCU Program of
National Research Foundation of Korea under Contract
No. R32-2008-000-10155-0.

\end{document}

%% file: authors_PUB.tex
\author{ M.~Ablikim$^{1}$, M.~N.~Achasov$^{9,e}$, S. ~Ahmed$^{14}$,
  X.~C.~Ai$^{1}$, O.~Albayrak$^{5}$, M.~Albrecht$^{4}$,
  D.~J.~Ambrose$^{44}$, A.~Amoroso$^{49A,49C}$, F.~F.~An$^{1}$,
  Q.~An$^{46,a}$
% ...
  \\
  \vspace{0.2cm}
  (BESIII Collaboration)\\
  \vspace{0.2cm} {\it
    $^{1}$ Institute of High Energy Physics, Beijing 100049, People's Republic of China\\
% ...
$^{4}$ Bochum Ruhr-University, D-44780 Bochum, Germany\\
    $^{5}$ Carnegie Mellon University, Pittsburgh, Pennsylvania 15213, USA\\
% ...
$^{9}$ G.I. Budker Institute of Nuclear Physics SB RAS (BINP), Novosibirsk 630090, Russia\\
% ...
$^{14}$ Helmholtz Institute Mainz, Johann-Joachim-Becher-Weg 45, D-55099 Mainz, Germany\\
% ...
$^{44}$ University of Rochester, Rochester, New York 14627, USA\\
% ...
$^{46}$ University of Science and Technology of China, Hefei 230026, People's Republic of China\\
% ...
$^{49}$ (A)University of Turin, I-10125, Turin, Italy; (B)University of Eastern Piedmont, I-15121, Alessandria, Italy; (C)INFN, I-10125, Turin, Italy\\
% ...
$^{a}$ Also at State Key Laboratory of Particle Detection and Electronics, Beijing 100049, Hefei 230026, People's Republic of China\\
% ...
$^{e}$ Also at the Novosibirsk State University, Novosibirsk, 630090, Russia\\
% ...
  }
}

%% file: Draft_etapri_jpsi.bbl
\begin{thebibliography}{99}

\bibitem{PDG}
K.~A.~Olive {\it et al.} [Particle Data Group], Chin.\ Phys.\ C
{\bf 38}, 090001 (2014).

\bibitem{PM}
T.~Barnes, S.~Godfrey, and E.~S.~Swanson,
Phys.\ Rev.\ D {\bf 72}, 054026 (2005).

\bibitem{Y(4260)1}
B.~Aubert {\it et al.}  [BaBar
 Collaboration], Phys.\ Rev.\ Lett.\
{\bf 95}, 142001 (2005); J.~P.~Lees {\it et al.} [BABAR
Collaboration], Phys.\ Rev.\ D\ {\bf 86}, 051102(R) (2012).

 \bibitem{Y(4360)1}
B.~Aubert {\it et al.}  [BABAR Collaboration], Phys.\ Rev.\ Lett.\
{\bf 98}, 212001 (2007); J.~P.~Lees {\it et al.} [BABAR
Collaboration], Phys.\ Rev.\ D\ {\bf 89}, 111103(R) (2014).

 \bibitem{Y(4260)2}
C.~Z.~Yuan {\it et al.} [Belle Collaboration], Phys.\ Rev.\ Lett.\
{\bf 99}, 182004 (2007); Z.~Q.~Liu  {\it et al.} [Belle
Collaboration], Phys.\ Rev.\ Lett.\ {\bf 110}, 252002 (2013).

\bibitem{Y(4360)2}
X.~L.~Wang {\it et al.} [Belle Collaboration], Phys.\ Rev.\ Lett.\
{\bf 99}, 142002 (2007); G.~Pakhlova {\it et al.} [Belle
Collaboration], Phys.\ Rev.\ Lett.\ {\bf 111}, 172001 (2008);
X.~L.~Wang {\it et al.} [Belle Collaboration], Phys.\ Rev.\ D {\bf
91}, no. 11, 112007 (2015).

\bibitem{Y(4260)3}
Q.~He {\it et al.} [CLEO Collaboration], Phys.\ Rev.\ D\ {\bf 74},
091104(R) (2006).

\bibitem{epjc_review} For a recent review, see N. Brambilla {\em et al.},
Eur. Phys. J. C {\bf 71}, 1534 (2011).

\bibitem{Coan:2006rv}
T.~E.~Coan {\it et al.}  [CLEO Collaboration], Phys.\ Rev.\ Lett.\
{\bf 96}, 162003 (2006).

\bibitem{Wang:2012bgc}
X.~L.~Wang {\it et al.}  [Belle Collaboration],  Phys.\ Rev.\ D
{\bf 87}, 051101 (2013).

\bibitem{Ablikim:2012ht}
M.~Ablikim {\it et al.} [BESIII Collaboration], Phys.\ Rev.\ D
{\bf 86}, 071101 (2012);  M.~Ablikim {\it et al.} [BESIII
Collaboration], Phys.\ Rev.\ D {\bf 91}, 112005 (2015).

\bibitem{Wang:2011yh}
Q.~Wang, X.~-H.~Liu and Q.~Zhao, Phys.\ Rev.\ D {\bf 84}, 014007
(2011).

\bibitem{qiao:2014}
C.~F.~Qiao, R.~L.~Zhu, Phys.\ Rev.\ D {\bf 89}, 074006 (2014).

\bibitem{ecm_gaoq}
M.~Ablikim {\it et al.} [BESIII Collaboration], Chin.\ Phys.\ C
{\bf 40}, 063001 (2016).

\bibitem{ref:bes3}
M.~Ablikim {\it et al.} [BESIII Collaboration], Nucl.\ Instrum.\
Meth.\ A {\bf 614}, 345 (2010).

\bibitem{ref:luminosity}
M.~Ablikim {\it et al.} [BESIII Collaboration], Chin.\ Phys.\ C
{\bf 39}, 093001 (2015).

\bibitem{ref:bes2}
J.~Z.~Bai {\it et al.} [BES Collaboration], Nucl.\ Instrum.\
Meth.\ A {\bf 458}, 627~(2001); {\bf 344}, 319 (1994).

\bibitem{ref:bes3physics}
D.~M.~Asner {\it et al.}, Int.\ J.\ Mod.\ Phys.\ A {\bf 24}, S1 (2009).

\bibitem{geant4}
S.~Agostinelli {\it et al.} [GEANT4 Collaboration], Nucl.\
Instrum.\ Meth.\ A {\bf 506}, 250 (2003).

\bibitem{kkmc}
S.~Jadach, B.~F.~L.~Ward and Z.~Was, Phys.\ Rev.\ D {\bf 63},
113009 (2001).

\bibitem{event}
R.~G.~Ping, Chin.\ Phys.\ C {\bf 32}, 599 (2008);
D.~J.~Lange, Nucl.\ Instrum.\ Meth.\ A {\bf 462}, 152 (2001).

\bibitem{lundcharm}
J.~C.~Chen, G.~S.~Huang, X.~R.~Qi, D.~H.~Zhang and Y.~S.~Zhu,
  Phys.\ Rev.\ D {\bf 62}, 034003 (2000).

\bibitem{ref:PYTHIA}
http://home.thep.lu.se/$\sim$torbjorn/Pythia.html

\bibitem{bayes}
J.~Conrad {\it et al.}, Phys.\ Rev.\ D {\bf 67}, 012002 (2003).

\bibitem{VP}
S.~Actis {\it et al.}  [Working Group on Radiative Corrections and
Monte Carlo Generators for Low Energies Collaboration], Eur.\
Phys.\ J.\ C {\bf 66}, 585 (2010).

\bibitem{QED}
E.~A.~Kuraev and V.~S.~Fadin, Sov.\ J.\ Nucl.\ Phys.\  {\bf 41},
466 (1985) [Yad.\ Fiz.\ {\bf 41}, 733 (1985)].

\bibitem{photon}
M.~Ablikim {\it et al.} [BESIII Collaboration], Phys.\ Rev.\ D
{\bf 81}, 052005 (2010).

\bibitem{helix}
M.~Ablikim {\it et al.} [BESIII Collaboration], Phys.\ Rev.\ D
{\bf 87}, 012002 (2013).

\end{thebibliography}
